\documentstyle[12pt,twoside,fleqn,espcrc1]{article}

\newcommand{\AmS}{{\protect\the\textfont2
  A\kern-.1667em\lower.5ex\hbox{M}\kern-.125emS}}
\input epsf.sty
\input psfig.sty
\title {Fission fragment mass reconstruction from Si surface
barrier detector measurement} 

\author{J.Velkovska\address{Department of Physics and Astronomy,
        State University of New York at Stony Brook, Stony Brook, 
	NY-11794, USA} %
    \thanks{corresponding author e-mail:
    julia@skipper.physics.sunysb.edu ,
	phone: (516) 632 3273 .
	This work was funded in part by the United States National
        Science Foundation.}
	and R.L. McGrath}
 
\begin{document}

\maketitle

Pacs:29.40-n,29.40Wk

\begin{abstract}
A method for plasma delay and pulse-height defect corrections for Si surface
barrier detectors (SBD) is presented. Based on known empirical
formulae, simple approximations involving the {\it measured}
time-of-flight (TOF) and energy of the ions were found and a mass 
reconstruction procedure was developed.
The procedure was applied for obtaining 
the fission  fragment mass and  angular distributions from the
$^{64}$ Ni+$^{197}$Au reaction at 418~MeV  and 383~MeV incident energy
using an array of eight SBDs. 
 
\end{abstract}
\newpage
  
\begin{figure}
\begin{center}
\centerline{\psfig{file=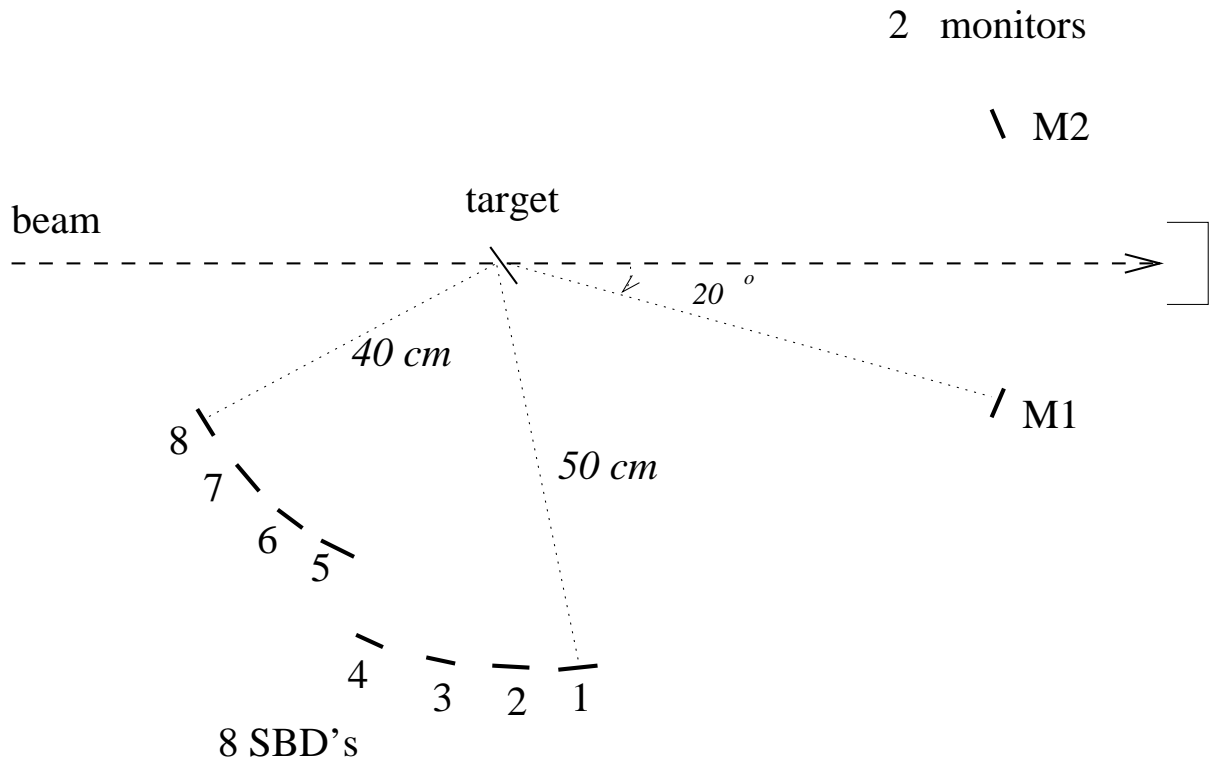,width=16cm,angle=0}}
\end{center}
\caption{ Experimental set-up used for fragment mass and angular
distribution  measurement from   the $^{64}$  Ni+$^{197}$Au reaction at
418~MeV and 383~MeV incident energy. Eight SBDs model EG\&G ORTEC
BF-028-400-60 were placed on a movable platform 10$^{\rm o}$
apart. Measurements were taken at two positions of the platform
covering from 20$^{\rm o}$ to 90$^{\rm o}$ and from 80$^{o}$ to
150$^{o}$ in the laboratory system.
\label{f:fig1}}
\end{figure}

Time-of-flight and energy measurements are the standard tool for the 
mass reconstruction of heavy-ion reaction products.
Si detectors can provide both of these quantities for ions that are
completely stopped in the detector material. A relatively simple
set-up (Fig.~\ref{f:fig1}) utilizing eight SBDs located at distances
50 cm and 40 cm from the target on a movable platform was used for the
fragment mass and angular distribution measurement from the 
 $^{64}$ Ni+$^{197}$Au reaction at 418~MeV and 383~MeV incident 
energy~\cite{PRC98}. The experiment was performed at the Stony Brook
Nuclear Structure Laboratory, using $^{64}$Ni beam from the FN Tandem
Van de Graaff and Superconducting LINAC accelerators. 

However, when SBDs are used for the detection of heavy energetic
fission fragments, the absolute flight-time and incident energy of the
ions can not
be directly determined from the measurement due to the plasma delay time
$\tau_{d}$ and the pulse-height defect $E_{phd}$. The heavy ions
impinging on the detector create a dense cloud of electron-hole pairs,
which can not be penetrated by the electric field generated by the
bias voltage, until the cloud is dispersed by the charge diffusion.
As a result, the charge collection is delayed by $\tau_{d}$ with
respect to the electron-hole creation time (which can be considered 
instantaneous). Within the dense electron-hole cloud, charge carriers
can recombine with consequent loss of pulse amplitude $E_{phd}$, hence
the total charge collected from the detector is no longer proportional
to the ion's energy. For the reactions studied above, $\tau_{d}$
was between 0.2 - 1.5 $ns$ and $E_{phd}$ - up to 25~MeV. Large
corrections to the measured TOF and energy $E_{det}$ were needed for
the successful mass reconstruction.

Although systematic studies of the above
effects exist in the literature \cite{Boh85,Ogi86}, the empirical
formulae that are
derived relate $\tau_{d}$ and $E_{phd}$ to the {\it unknown} mass A, charge Z
and incident energy $E=E_{det}+E_{phd}$ of the ions:
\begin{equation}
\label{eq:td}
\tau_{d}=\frac{\overline{Z}^{2}}{2A}\exp\left(\frac{-E}{3.75A}\right)C
\left(\frac{\rho}{d}\right)^{\frac{2}{5}}\frac{1}{F_{eff}} 	,
\end{equation}
\begin{equation}
\label{eq:phd}
E_{phd}=2.33\times
10^{-4}(AZ)^{\frac{6}{5}}(E/A)^{\frac{1}{2}}\times\left[1+{{1.32\times
10^{3}Z(S/E^{2})^{\frac{1}{3}}}\over{\rho^{\frac{1}{4}}F_{eff}}}\right]
\end{equation}   
Here, $\overline{Z}\approx\sqrt{\left(dE/dx\right)_{ion}/\left(dE/dx\right)_{proton}
}$ (charge units) is the effective charge of the ion, expressed in
terms of the stopping power of the ion and a proton at the same
velocity, 
d~(cm) - the detector thickness,
$\rho$ ($\Omega\cdot cm$) - the detector resistivity,
C (pF) - the capacitance of the totally depleted detector, S - the
stopping power in the detector material in MeV/(mg/cm$^{2}$).
The effective field strength $F_{eff}$ (V$\cdot$cm) is given by:
\begin{equation}
F_{eff}(x)=(d-x)/\mu\tau	,
\end{equation}
where $d\approx \sqrt{2\tau\mu V}$ is the depletion depth at a given bias
voltage $V$, $\mu\approx~1481$ $cm^{2}V^{-1}s^{-1}$ is the constant electron
mobility, $\tau\approx \rho\times 10^{-12} s$ and $x$ is taken at the charge 
centroid of the ionized track produced by the heavy ion (about 1/3 of the ion 
range in the material).
Some relevant quantities for the detectors used in this experiment are given
in Table~\ref{tab:table1}.

Being  able  to calculate  $\tau_{d}$  and  $E_{phd}$  from the  above
equations does not   guarantee that the  necessary  corrections to the
measured  TOF  and  $E_{det}$  can be   made, since   finding analytic
solutions to the  inverse problems is  extremely complicated.  In this
work, approximate solutions  which are simple,  accurate to  $5\%$ and
derived directly from the measured quantities are described. 

\begin{figure}
\begin{center}
\centerline{\psfig{file=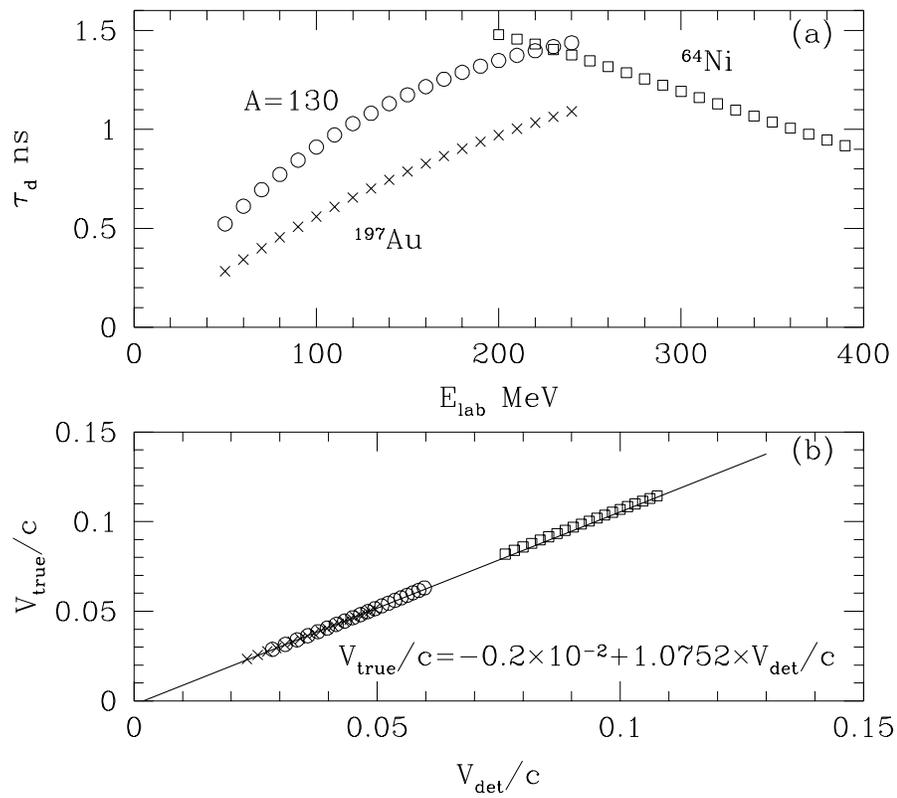,height=12cm,angle=0}}
\end{center}
\caption{(a) Plasma delay time for selected ions as a function of
incident energy. (b) The true velocity $V_{true}$ as a function of the
experimentally accessible velocity $V_{det}$ with the fitted linear 
expression.\label{f:nim1}}
\end{figure}

The  plasma  delay  time was calculated from equation~(\ref{eq:td})   
for ions with masses and energies  within the range of  the
experiment. Figure~\ref{f:nim1}(a) shows  the delay
time for the  projectile $^{64}$Ni, the  recoil $^{197}$Au and for the
symmetric fragments A=130 as a function of incident energy. There is
no unique functional dependence on the incident energy for the
different ions of interest, hence the energy information
from the measurement can not be used to correct the measured TOF.
However, if we plot the ``true'' velocity  of the ions as a  function
of the measured velocity 
$V_{det}=flight\_distance/(TOF+\tau_{d})$~(Fig.\ref{f:nim1}(b)), for
all ions in the  range of the experiment,  there is one linear function giving
the relation  between the  ``true'' and  the  measured velocity.  The
fitted line and the fit parameters are shown in
Fig.\ref{f:nim1}(b). Thus using equation~(\ref{eq:td}) and the
geometry of the particular experiment, one can extract a linear
correction to the measured velocity.
 
\begin{figure}
\begin{center}
\centerline{\psfig{file=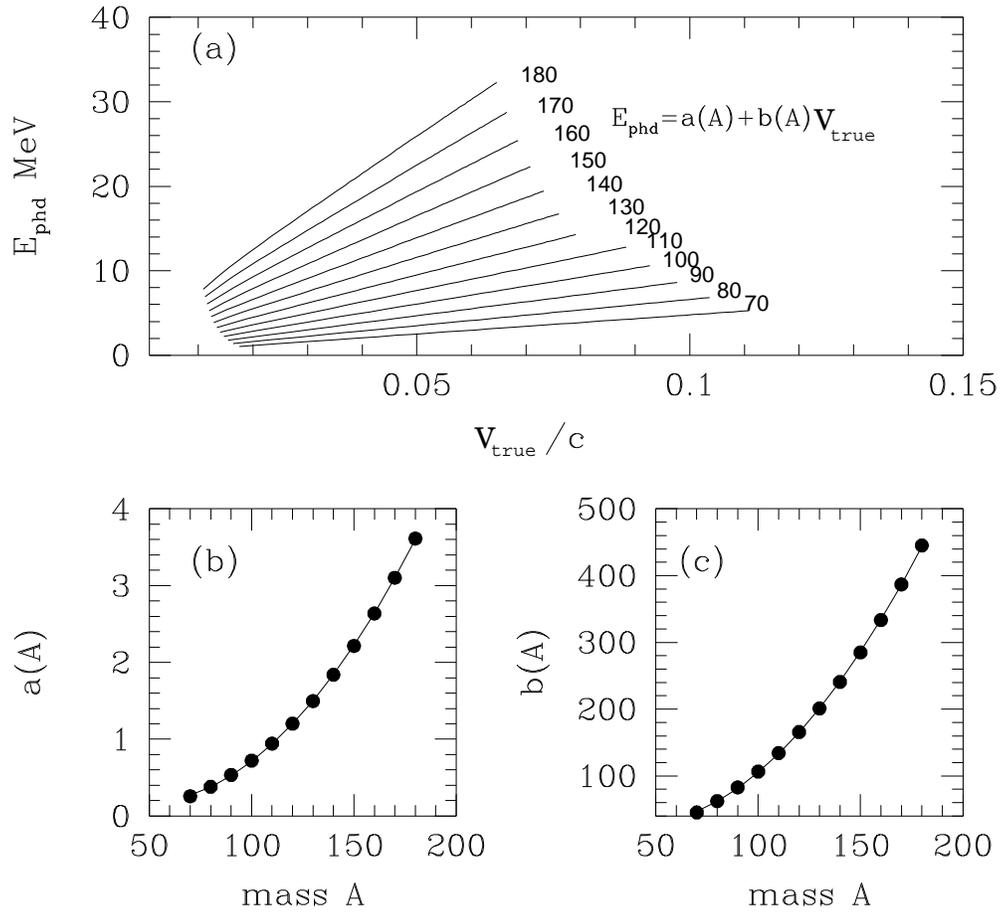,height=12cm,angle=0}}
\end{center}
\caption{(a) Constant mass curves of pulse-height defect as a function of
velocity. (b),(c) The coefficients of the linear fits (points)
$E_{phd}(V_{true})$ and the quadratic fits to their mass dependence
(lines).\label{f:nim2}}
\end{figure}

After applying the velocity corrections, we need to account for the 
pulse-height defect in order to reconstruct the ion mass using 
equation~(\ref{eq:phd}) and the measured energy $E_{det}=E-E_{phd}$.
Although deriving an equation for the mass A involving the corrected
velocity and other known quantities is in principle possible, its
analytic solution is a formidable task. The following simple procedure
was developed instead: 

The pulse-height defect was calculated (from equation~(\ref{eq:phd}))
for the masses and energies covered in the range of the
experiment. Assuming that the charge-to-mass ratio in the fragments is equal
to the charge-to-mass ratio in the composite system, the
dependence on the velocity for constant mass was extracted. 
Figure~\ref{f:nim2} shows the constant mass curves: they show linear 
dependence of $E_{phd}$ on the velocity.
Only for $V_{true}<0.02c$ a deviation from linearity is observed. This
imposed a velocity cut in the data analysis.
Linear fits were applied to the calculated constant mass curves: 
$E_{phd}(V_{true})=a(A)+b(A)\cdot V_{true}$. The coefficients a(A) and
b(A) at fixed mass were extracted for A=70-180 in steps of 10 mass units.

After factoring out the velocity dependence, the mass dependence of the
coefficients a(A) and b(A) was examined. They were found to depend 
quadratically on mass. The quadratic fits are shown in panel (b) and (c) of
Figure~\ref{f:nim2}. 
With the parameters extracted from the above fits, we can write a quadratic 
equation for the mass of the ion, involving only {\it measured} quantities:

\begin{equation}
\label{eq:phd1}
E_{det}+\left[a_{1}+a_{2}A+a_{3}A^{2}+\left(b_{1}+b_{2}A+b_{3}A^{2}
\right)V_{true}\right]={AV_{true}^{2}u\over 2}
\end{equation}
Here $V_{true}$ is the ``true'' velocity, which was previously determined as a
function of the measured velocity $V_{det}$ and $u=931.5$~MeV is     
the nucleon  mass. 
 
A computer simulation was developed to test this procedure. Fission
fragments from the  $^{64}$ Ni+$^{197}$Au reaction at 418~MeV were
generated with a known mass A and Viola systematics kinetic energy. 
The laboratory energy and TOF were calculated for each detector
position. Using equations~(\ref{eq:td}) and~(\ref{eq:phd}) the plasma delay
and the pulse-height defect were calculated. Then $E_{det}$ and
$V_{det}$ were obtained, and following the outlined procedure, the
``unknown'' mass of the fragments was recovered. 
The  solution of
equation~(\ref{eq:phd1})  gave very  accurate  results for   light masses
(within  1.5 mass units) and  somewhat poorer results for the heaviest
masses  (within   5 units  of the   input). 

\begin{figure}
\begin{center}
\centerline{\psfig{file=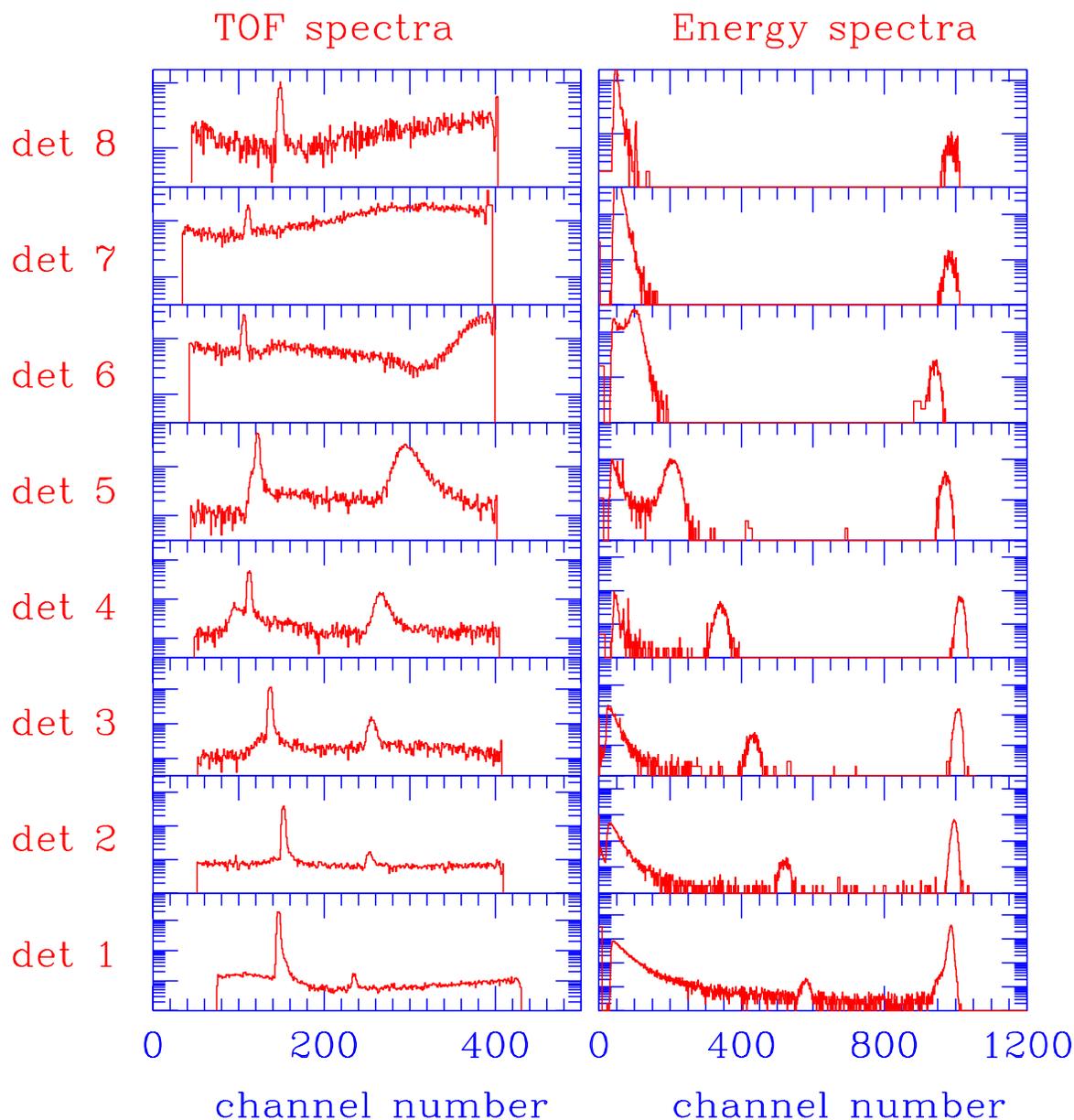,height=17cm,angle=0}}
\end{center}

\caption{ TOF and energy spectra in the fission-fragment detectors from the
247 MeV calibration run. The sharp peaks in the TOF spectra come from the
elastically scattered $^{64}$Ni. The broad peaks to the right are the recoil
$^{197}$Au. In the energy spectra, the elastic $^{64}$Ni is at channel$\sim
1000$. The lower peaks come from $^{197}$Au.\label{f:nim3}}
\end{figure}

Another technical detail, that is crucial for this measurement is the
absolute energy and time calibration of the detectors and the
subsequent electronics. 
The energy calibration was done using several reference signals:
The low energy part of the spectrum was calibrated using a $^{228}$Th
$\alpha$-~source.
Detectors 1-5 in the forward position (covering from 20$^{o}$ to
90$^{o}$ in the laboratory system) were within the grazing angle for the 
$^{64}$Ni + $^{197}$Au,~E$_{lab}=418$~MeV reaction, so the elastic peak could
be used as a calibration point. A low energy calibration run at  
247~MeV provided additional calibration points. At
this energy, with the platform in forward position, all SBDs could
detect Rutherford scattering. Fig.~\ref{f:nim3}
shows the TOF and the energy spectra in all eight detectors at this energy.
It is interesting to note, that in SBD 1-5 the recoil $^{197}$Au is also
seen. These points, wherever present, were included in the calibration 
as well. The energy resolution was $\sim2\%$ for the elastically
scattered beam and $\sim1.2\%$ for $\alpha$-particles. 

To calibrate the timing one needs to relate the channel number of the TOF 
spectrum to the actual TOF through a linear function. The slope was determined 
reliably using the reference RF signal of the LINAC accelerator. 
Determining the absolute offset is more difficult, because at the high energy
only the first 5 detectors could see elastically scattered beam, for which the
TOF is known absolutely. The 247~MeV run gave the relative timing between the
detectors, so SBDs 6,7,8 could also be calibrated.
The Full-Width-at-Half-Maximum of the  beam pulse coming from the
accelerator was $\le~600ps$. Corrections were made for the plasma delay time of
the heavy fragments, and the overall timing resolution for the elastic reaction
products measured was $\le~1 ns$.

\begin{figure}
\begin{center}
\centerline{\psfig{file=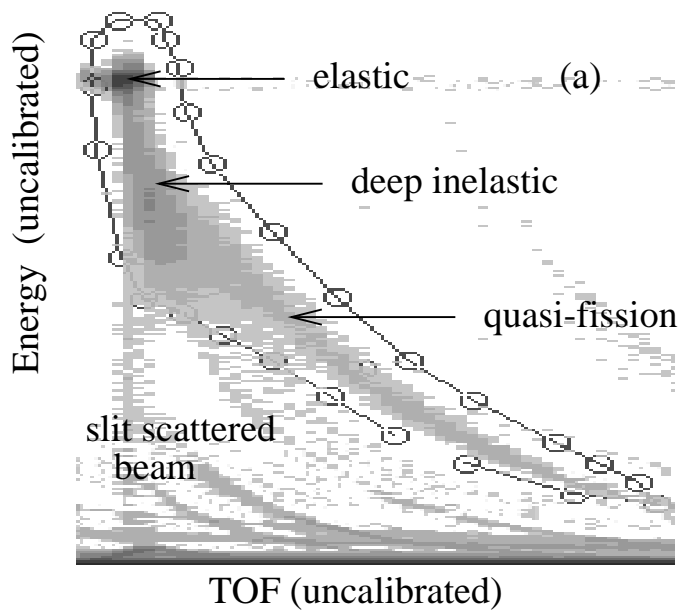,height=8cm,angle=0}}
\end{center}
\begin{center}
\centerline{\psfig{file=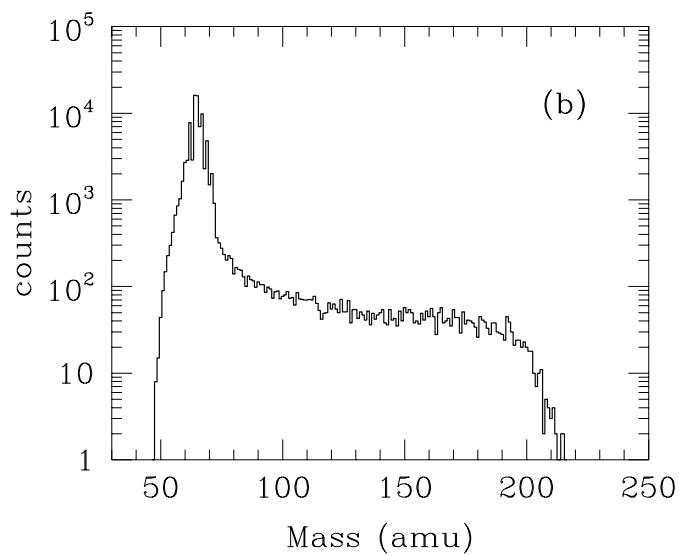,height=9cm,angle=0}}
\end{center}
\vskip -1 cm
\caption{(a) Energy versus TOF 
(uncalibrated) scatter plot from detector 3 located at 40$^{\rm o}$ in the
lab system for E$_{lab}=418$~MeV. The ``banana'' gate is applied to select the 
reaction products of interest. (b) Mass reconstruction for the gated
events. The peak in the spectrum is the elastically scattered $^{64}$Ni. 
\label{f:mass}}
\end{figure}

After completing the time and energy calibrations, we have all the
ingredients needed for the mass reconstruction procedure. One should
note, that it has built-in the assumption that the reaction products
are coming from the target (which is at a known distance from the
detector), so care should be taken that noise events do not enter the
reconstruction. Figure~\ref{f:mass}(a) shows an example of a scatter
plot of uncalibrated energy versus TOF from detector 3, located at
40${^o}$ in the lab, from the 418 MeV measurement. The reaction
products of interest are separated from the slit scattered beam and
low pulse-height noise using a ``banana'' gate. Only the gated events
are reconstructed. Figure~\ref{f:mass}(b) shows the reconstructed mass
spectrum. Combining all experimental uncertainties, the mass
reconstruction was  estimated to be accurate to $\leq 10$ mass units.

\bibliographystyle{unsrt}
\bibliography{bib}

\begin{thebibliography}{1}

\bibitem{PRC98}
J.~Velkovska, C.~R. Morton, R.~L. McGrath, P.~Chung, and I.~Di\'{o}szegi.
\newblock {\em Phys. Rev.}, C 59:1518, 1999.

\bibitem{Boh85}
W.~Bohne, W.~Galster, K.~Grabishc, and H.~Morgenstern.
\newblock {\em Nucl. Instr. and Meth.}, A240:145, 1985.

\bibitem{Ogi86}
M.~Ogihara, Y.~Nagashima, W.~Galster, and T.~Mikumo.
\newblock {\em Nucl. Instr. and Meth.}, A251:313, 1986.

\end{thebibliography}
\newpage

\begin{table}[hbt]
\newlength{\digitwidth} \settowidth{\digitwidth}{\rm 0}
\catcode`?=\active \def?{\kern\digitwidth}
\caption{BF-028-400-60 detector characteristics}
\label{tab:table1}

\begin{tabular*}{\textwidth}{@{}l@{\extracolsep{\fill}}rr}
\hline

Detector Model  &         Nominal            \\
BF-028-400-60   &      Characteristics        \\

\hline

Active area [mm$^{2}$  		&	400\\
Resistivity [$\Omega\cdot cm$]  &	490\\
Thickness [cm]			&       0.006\\
Capacitance [pF]		&  	163\\
Nominal bias voltage [V]	&   	64\\
Effective field strength 	&		\\
at nominal bias  [V/cm]	 	&	12362\\
Bias voltage applied [V]	&   	100\\
Effective field strength 	&	\\
at 100 V  bias  [V/cm]	 	& 15680	\\

\hline
\end{tabular*}
\end{table}

\newpage

\end{document}